\documentclass[twocolumn,prl,showpacs,preprintnumbers,amsmath,amssymb]{revtex4-1}

\usepackage{graphicx}% Include figure files \usepackage{dcolumn}
\begin{document}

%\preprint{APS/123-QED}

\title{Coexisting pseudo-gap and superconducting gap in the high-$T_c$ superconductor La$_{2-x}$Sr$_x$CuO$_4$}

\author{T. Yoshida$^1$, W. Malaeb$^1$, S. Ideta$^1$, A. Fujimori$^1$, D. H. Lu$^2$, R. G. Moor$^2$, Z.-X. Shen$^2$, M. Okawa$^3$, T. Kiss$^3$, K. Ishizaka$^3$, S. Shin$^3$, Seiki Komiya$^4$, Yoichi Ando$^5$, H. Eisaki$^6$, S. Uchida$^1$}

\affiliation{$^1$Department of Physics, University of Tokyo,
Bunkyo-ku, Tokyo 113-0033, Japan}

\affiliation{$^2$Department of Applied Physics and Stanford
Synchrotron Radiation Laboratory, Stanford University, Stanford,
CA94305}

\affiliation{$^3$Institute of Solid State Physics, University of
Tokyo, Kashiwa 277-8581, Japan}

\affiliation{$^4$Central Research Institute of Electric Power
Industry, Komae, Tokyo 201-8511, Japan}

\affiliation{$^5$Institute of Scientific and Industrial Research,
Osaka University, Ibaraki, Osaka 567-0047, Japan}

\affiliation{$^6$National Institute of Advanced Industrial Science
and Technology, Tsukuba 305-8568, Japan}

\date{\today}

\begin{abstract}
Relationship between the superconducting gap and the pseudogap has
been the subject of controversies. In order to clarify this issue,
we have studied the superconducting gap and pseudogap of the
high-$T_c$ superconductor La$_{2-x}$Sr$_x$CuO$_4$ ($x$=0.10, 14)
by angle-resolved photoemission spectroscopy (ARPES). Through the
analysis of the ARPES spectra above and below $T_c$, we have
identified a superconducting coherence peak even in the anti-nodal
region on top of the pseudogap of a larger energy scale. The
superconducting peak energy nearly follows the pure $d$-wave form.
The $d$-wave order parameter $\Delta_0$ [defined by
$\Delta(k)=\Delta_0(\cos k_xa-\cos k_ya)$ ] for $x$=0.10 and 0.14
are nearly the same, $\Delta_0\sim$12-14 meV, leading to strong
coupling $2\Delta_0/k_BT_c\sim 10$. The present result indicates
that the pseudogap and the superconducting gap are distinct
phenomena and can be described by the ``two-gap" scenario.
\end{abstract}

% Classification Scheme. %\keywords{Suggested keywords}%Use show keys class option if keyword

\maketitle

%Pseudogap issue
In the studies of the high-$T_c$ cuprates, it has been a
long-standing issue whether the pseudogap is related to the
superconductivity or a phenomenon distinct from the
superconductivity. Preformed Cooper pairs lacking phase coherence
\cite{Kivelson} or superconducting fluctuations \cite{Engelbrecht}
have been proposed as a possible origin of the pseudogap.
Alternatively, the pseudogap is attributed to a competing order
such as spin density wave, charge density wave, loop current
\cite{Varma} etc. In measurements which are sensitive to the
superconducting gap around the node such as Andreev reflection,
penetration depth, and Raman scattering, the gap decreases with
underdoping \cite{Deustcher,Panagopoulos,Opel,Tacon} in contrast
to the pseudo-gap which increases with underdoping, suggesting a
different origin of the antinodal gap from the superconducting
gap. Furthermore, recent ARPES studies have indicated the presence
of two distinct energy scales as well as distinct momentum and
temperature dependences of the gaps in the nodal and anti-nodal
regions \cite{Tanaka,Lee,Kondo,Terashima,Hashimoto}. In a previous
paper \cite{Yoshida}, we have pointed out that the pseudo-gap
shows a relatively material-independent universal behavior: the
pseudo-gap size is almost the same at the same doping level while
the superconducting gap is proportional to $T_c$, suggesting
different origins for the superconducting gap and the pseudo-gap.
On the other hand, a simple $d$-wave-like gap has been also
reported in some ARPES studies \cite{Meng,Shi}. In such a
single-gap picture, the pseudogap is interpreted as a signature of
preformed Cooper pairs. Thus, the discrepancy between the
experimental studies has remained.

%Present study
In the analysis of the STM spectra of single-layer cuprate
Bi$_2$Sr$_2$CuO$_{6+\delta}$ (Bi2201), distinct behaviors of the
superconducting gap and the pseudogap have been clearly
demonstrated \cite{Boyer}. Even if the superconducting peak is not
clearly observed in underdoped samples, the superconducting
coherence peak has been identified by dividing the spectra in the
superconducting state by the normal-state data. Also, similar
analysis has been done for the ARPES spectra of Bi2201 in the
anti-nodal region and a superconducting peak has been identified
\cite{Ma}. These results suggest that the pseudogap and the
superconducting gap have distinct origins and the superconducting
gap is created on top of the electronic states with relatively
broad spectral features with low density of states due to the
pseudogap opening. As for the single-layer cuprates
La$_{2-x}$Sr$_x$CuO$_4$ (LSCO), a clear superconducting peak has
been identified in the off-nodal region, however, such a clear
peak has not been identified in the anti-node region
\cite{Terashima}. In order to examine the coexistence of the
superconducting gap and the pseudogap aforementioned, we have
performed an ARPES study of LSCO ($x$=0.14, 0.10) and analyzed the
spectral line shapes to extract the signature of the
superconducting peak.

%Experiment
High-quality single crystals of LSCO ($x$=0.10, 0.14, 0.15) were
grown by the traveling- solvent floating-zone method. The critical
temperatures of $T_c$'s the $x$ = 0.10, 0.14, and 0.15 samples
were 28, 32 and 39 K, respectively. The ARPES measurements were
carried out using synchrotron radiation at beamline 5-4 of
Stanford Synchrotron Radiation Light Source (SSRL) and also using
a UV laser at the Institute of Solid State Physics (ISSP), the
University of Tokyo. We used incident photons with energies of 22
eV at SSRL and 6.994 eV at ISSP. SCIENTA R4000 spectrometers were
used in the angle mode. The total energy resolution was about 7
meV at SSRL and 2.8 meV at ISSP. The samples were cleaved
\textit{in situ} and measurements were performed at 11 K or 4 K
($<T_c$) and 40 K ($>T_c$).

\begin{figure}
\includegraphics[width=8.5cm]{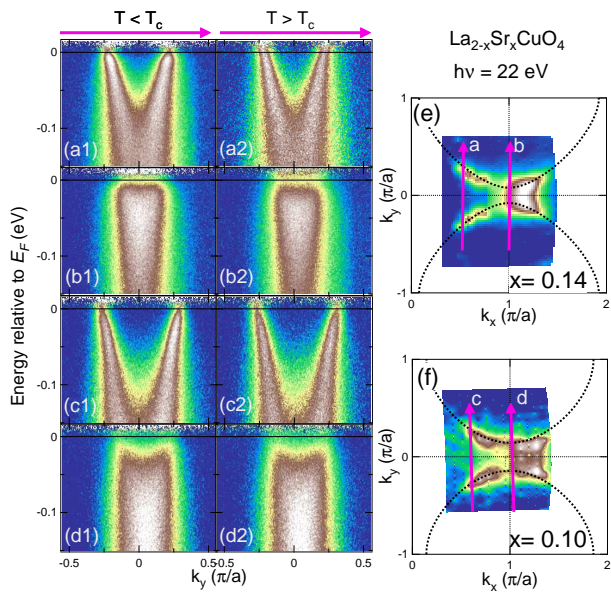}
\caption{\label{EkMap}(Color online) ARPES spectra of LSCO with
$x$=0.14 and 0.10. (a1)-(d2) ARPES intensity plots correspond to
cuts a-d in panels (e) and (f). The data in the left (right) have
been measured below (above) $T_c$. Spectra have been divided by
the Fermi-Dirac function convoluted with the energy resolution
function. (e)(f) Intensity at $E_F$ mapped in the $k_x$-$k_y$
plane. Dotted lines illustrate Fermi surfaces.}
\end{figure}

%Ekmap
Figure \ref{EkMap} shows ARPES spectra of LSCO with $x$=0.14 and
0.10 taken at $T$= 11K ($<T_c$) and $T$= 40K ($>T_c$). The spectra
have been divided by a convoluted Fermi-Dirac function. One can
clearly see that the superconducting gap for the $x$=0.14 sample
in the off-nodal region opens below $T_c$ [panel (a1)] and closes
above $T_c$ [panel (a2)]. The spectra for the $x$=0.10 sample also
shows a similar trend but the gap opens even above $T_c$ as shown
in panel (c2), suggestive of a pseudogap opening. In the anti-node
region, in contrast, spectral weight near $E_F$ is strongly
suppressed even above $T_c$ [(b2) and (d2)], indicating a
pseudogap behavior. The difference between the spectra below and
above $T_c$ is not apparent. From these data, we derive the energy
of the superconducting peak as described below.

\begin{figure}
\includegraphics[width=8.5cm]{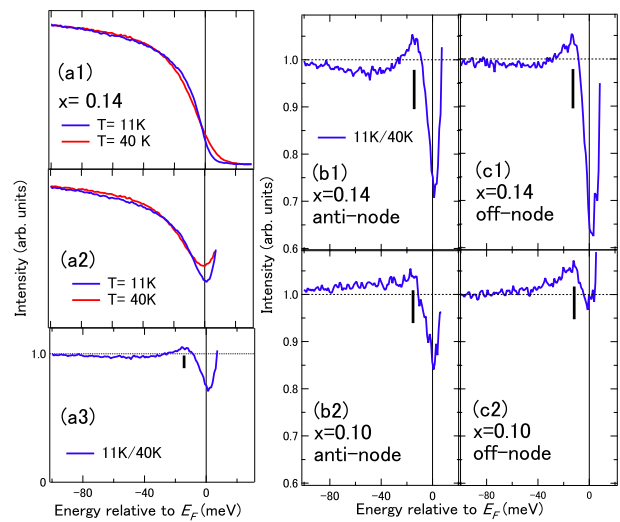}
\caption{\label{IMDC}(Color online) Superconducting peaks observed
in the ARPES spectra of LSCO with $x$=0.14 and 0.10. (a1)
Cut-integrated spectra for $x$=0.14 for cut b in Fig.\ref{EkMap}
(e). (a2) Spectra in panel (a1) have been divided by the Fermi
function convoluted with the energy resolution function. (a3)
Spectra below $T_c$ in panel (a2) have been divided by that above
$T_c$. (b1)-(c2) Spectra corresponding to cuts in Fig. \ref{EkMap}
after above processing. Vertical bars indicate the peak
positions.}
\end{figure}

%IMDC
In order to identify fine structures associated with the
superconducting transition, we have applied a similar analysis to
that employed in the previous STM \cite{Boyer} and ARPES studies
\cite{Ding} as described in Figs. \ref{IMDC} (a1) -(a3). First,
integrated spectrum along cut b in Fig. \ref{EkMap}[panel (a1)] is
divided by Fermi-Dirac function convoluted by the energy
resolution [panel (a2)]. Then, the spectrum below $T_c$ is divided
by that above $T_c$ [panel (a3)]. As a result, we have obtained a
peak-gap structure near $E_F$ even in the anti-nodal region where
the pseudo-gap dominates the spectra, indicating superconducting
peak and gap. In the same manner, the various cuts shown in Fig.
\ref{EkMap} have been analyzed and the results are shown in Fig.
\ref{IMDC} (b1)-(c2). Note that the obtained spectra is analogous
to the tunneling spectra of $s$-wave superconductors
\cite{McMillan} because the superconducting order parameter is
approximately constant around $k_F$ on a single cut.

\begin{figure*}
\includegraphics[width=17cm]{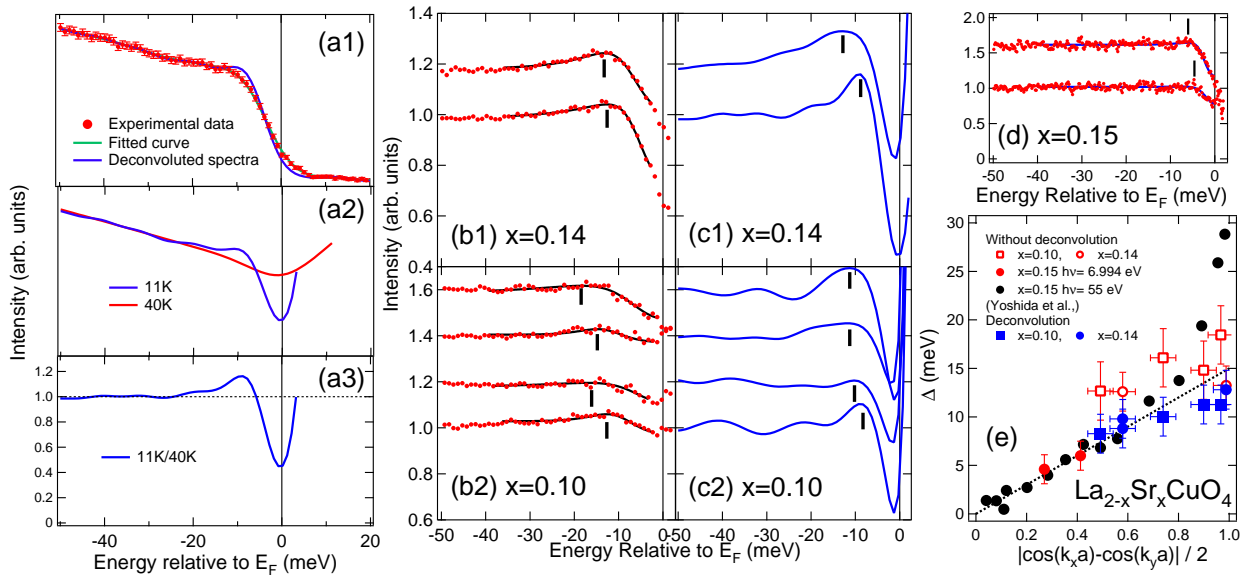}% Here is how to import EPS art
\caption{\label{Gap}(Color online) Angular dependence of the
superconducting peaks obtained from ARPES spectra of LSCO with
$x$=0.14 and 0.10. (a1) Cut-integrated spectrum for $x$=0.14 in
the off-nodal direction taken at 11 K and its deconvoluted
spectrum using the MEM. Fitted curve produced by the MEM well
reproduce the original experimental data. (a2) Deconvoluted
spectra divided by the Fermi-Dirac function. (a3) Spectrum
obtained by dividing the 11K data ($<T_c$) by the 40K data
($>T_c$)in panel (a2). (b1)(b2) Superconducting peak obtained in
Fig.\ref{IMDC} (c1)(c2). The spectra in panels (b1) and (b2) have
been deconvoluted with the energy resolution using the MEM. (d)
Superconducting gap near the nodal direction for $x$=0.15 taken at
$h\nu$=6.994 eV corresponding to panels (b1) and (b2). (e) Angular
dependence of the superconducting peak are plotted as a function
of $d$-wave parameter $|\cos(k_x)-\cos(k_y)|$/2. For comparison,
previous results of the gap for $x$=0.15 are also plotted. }
\end{figure*}

%MEM
Strictly speaking, the division by a convoluted Fermi-Dirac
function is an approximate method to determine the gap size and
one cannot exclude spurious effect due to the finite energy
resolution. In order to determine the superconducting gap energy
more precisely, we performed deconvolution to remove the
experimental energy resolution from the cut-integrated spectra
using the maximum entropy method (MEM) [Fig. \ref{Gap}(a1)]. Then,
the spectra were divided by the Fermi-Dirac function [Fig.
\ref{Gap}(a2)]. Finally, the spectra below $T_c$ were divided by
those above $T_c$ [Fig. \ref{Gap}(a3)]. In Fig. \ref{Gap}, we
compare the processed spectra with [panels (b1) and (b2)] and
without deconvolution [panels (c1) and (c2)]. Also, we have shown
processed spectra (without deconvolution) for $x$=0.15 near the
node taken by the laser ARPES with a high resolution of $\sim2.8$
meV [Fig. \ref{Gap}(d)]. The peak energies are plotted as a
function of the $d$-wave parameter $(\cos(k_x)-\cos(k_y))/2$ in
panel (e) and compared with the previous result of $x$=0.15 which
shows "two-gap" behavior \cite{Yoshida}. Note that the peak
positions of the deconvoluted spectra are closer to $E_F$ by
$\sim$ 5meV than those without deconvolution, nearly follows the
pure $d$-wave from the nodal to the anti-nodal regions
\cite{Yoshida}. Furthermore, the gap sizes in the off-nodal region
for both the $x$=0.14 and 0.10 samples are almost the same. The
observed $d$-wave-like gap in the anti-node region
$\Delta_0\sim$12-14 meV gives a strong coupling ratio
$2\Delta_0/k_BT_c\sim 10$, similar to the previous Bi2201 result
\cite{Ma}.

\begin{figure}
\includegraphics[width=8.5cm]{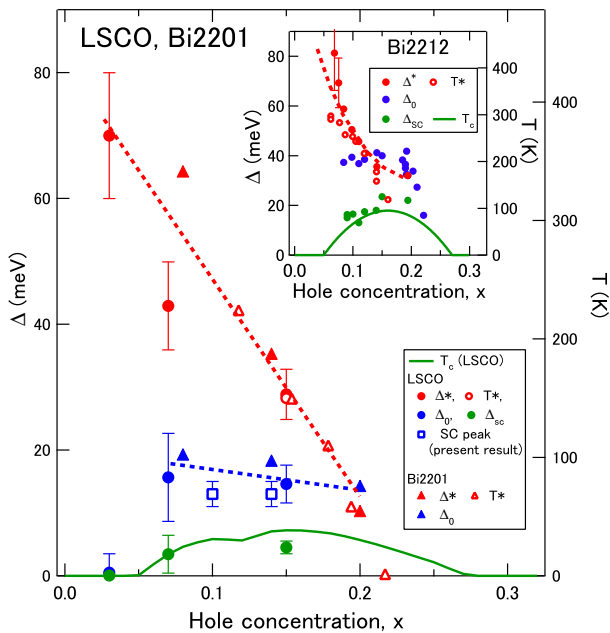}
\caption{\label{MaterialGap}(Color online) Doping dependences of
the characteristic energies
($\Delta^*$,$\Delta_0$,$\Delta_{sc}$\cite{Yoshida,Ideta,PGreview})
and temperatures ($T^*$, $T_c$) for the single-layer cuprates
(LSCO, Bi2201). The present LSCO result of the SC peak energy in
the antinode direction is also plotted. Inset shows those for the
double-layer cuprates Bi2212. Gap energies $\Delta$ and
temperatures $T$ have been scaled as $2\Delta = 4.3k_BT$ in these
plots. Parameter values have been taken from Ref.\cite{PGreview}
and references therein.}
\end{figure}

%Phenomenological description
Here, we shall discuss differences in the "two-gap" behavior
between the single-layer and bi-layer cuprates. In the
single-layer cuprates such as LSCO and Bi2201, the co-existence of
the pseudogap and the superconducting peak in the anti-nodal
direction has been revealed by the ARPES \cite{Ding} and STM
studies \cite{Kurosawa} including the present result. However, in
the case of Bi2212 \cite{Tanaka, Lee}, such two energy scales have
not been resolved in the anti-nodal spectra. Only a single-peak
structure appears below $T_c$ in the optimally-doped and
under-doped regions. The different behaviors between the single-
and double-layer cuprates can be understood as follows: When the
pseudogap has a different origin from the superconducting gap, the
superconducting peak is created on the pseudogap feature of the
broad incoherent spectral weight. In the single-layer cuprates,
which have relatively low $T_c$'s, the energy scale of the
superconducting gap is smaller than that of the pseudogap and the
$T_c$ is lower than the pseudogap temperature $T^*$ in the
optimally-doped and the underdoped region as shown in
Fig.\ref{MaterialGap}(a). Therefore, the superconducting gap
appears below $T_c$ within the pseudogap which is created below
$T^*$, resulting in the two energy scales in the spectral weight
distribution. On the other hand, in the bi-layer cuprates Bi2212,
which have a relatively high $T_c$ comparable to $T^*$, both
energy scales are comparable and $T_c$ and $T^*$ are also
comparable in the optimally doped region
[Fig.\ref{MaterialGap}(b)]. Because the pseudogap has a comparable
energy scale with the superconducting gap $\Delta_0$ and opens at
nearly the same temperatures, the two gaps cannot be clearly
resolved. Thus, the superconducting peak may show only a weak
deviation from the pure $d$-wave in Bi2212 \cite{Tanaka, Lee},
unlike the strong deviation with two-energy scales in the
single-layer cuprates \cite{Yoshida, Terashima, Kondo}.

%Microscopic model
The present study has revealed that the superconducting gap has
nearly pure $d$-wave form and exists even in the anti-node
direction in the optimally-doped to underdoped region. A
phenomenological model for the two gap state proposed by Yang,
Rice and Zhang (YRZ) well accords with the present coexistence of
the superconducting and the pseudogap \cite{YRZ}. While YRZ
assumes a RVB gap as the anti-nodal gap, there are several
possible candidates for the origin of the pseudogap. Calculations
assuming an order parameter different from the superconductivity
such as valence bond glass \cite{Ren} and spin-density wave (SDW)
state \cite{Das} have predicted that the superconducting gap
persists beyond the end of the Fermi arc all the way to the
antinode. In accordance with the present observation.
Particularly, in the SDW-based calculation \cite{Das}, a hump-like
pseudogap and a sharp superconducting peak in the anti-node
direction as seen in the present result have been reproduced. A
temperature dependent ARPES study reveals particle-hole asymmetry
of the anti-node gap, most likely due to the density-wave gap
formation \cite{HashimotoNatPhys}. This observation would be
related to the charge ordered state observed in STM
\cite{Hanaguri} or stripe formation \cite{Tranquada}. From a STM
result of the charge order \cite{Kurosawa}, the pseudogap in the
anti-nodal region is most likely to link to such a two-dimensional
electronic charge order.

%Conclusive remarks
In conclusion, we have identified the superconducting peak LSCO
($x$=0.10, 0.14) in the anti-nodal region from analysis of the
ARPES spectra above and below $T_c$. The superconducting peaks
follow the pure $d$-wave on top of the pseudogap of a larger
energy scalle. The $d$-wave gap parameter $\Delta_0$ of $x$=0.10
and 0.14 are nearly the same. Since the superconducting order
parameter is nearly doping independent in the underdoped region,
the drop of $T_c$ with underdoping is due to the decreasing length
of Fermi arc. The present results have reinforced that the
pseudogap and the superconducting gap are distinct phenomena.

%Acknowledgment
We are grateful to H. Ding for informative discussions. This work
was supported by a Grant-in-Aid for Young Scientists (B)(22740221)
and the Japan-China-Korea A3 Foresight Program from the Japan
Society for the Promotion of Science. SSRL is operated by the
Department of Energy's Office of Basic Energy Science, Division of
Chemical Sciences and Material Sciences.

\bibliography{SCpeak}

\begin{thebibliography}{28}%
\makeatletter
\providecommand \@ifxundefined [1]{%
 \@ifx{#1\undefined}
}%
\providecommand \@ifnum [1]{%
 \ifnum #1\expandafter \@firstoftwo
 \else \expandafter \@secondoftwo
 \fi
}%
\providecommand \@ifx [1]{%
 \ifx #1\expandafter \@firstoftwo
 \else \expandafter \@secondoftwo
 \fi
}%
\providecommand \natexlab [1]{#1}%
\providecommand \enquote  [1]{``#1''}%
\providecommand \bibnamefont  [1]{#1}%
\providecommand \bibfnamefont [1]{#1}%
\providecommand \citenamefont [1]{#1}%
\providecommand \href@noop [0]{\@secondoftwo}%
\providecommand \href [0]{\begingroup \@sanitize@url \@href}%
\providecommand \@href[1]{\@@startlink{#1}\@@href}%
\providecommand \@@href[1]{\endgroup#1\@@endlink}%
\providecommand \@sanitize@url [0]{\catcode `\\12\catcode `\$12\catcode
  `\&12\catcode `\#12\catcode `\^12\catcode `\_12\catcode `\%12\relax}%
\providecommand \@@startlink[1]{}%
\providecommand \@@endlink[0]{}%
\providecommand \url  [0]{\begingroup\@sanitize@url \@url }%
\providecommand \@url [1]{\endgroup\@href {#1}{\urlprefix }}%
\providecommand \urlprefix  [0]{URL }%
\providecommand \Eprint [0]{\href }%
\providecommand \doibase [0]{http://dx.doi.org/}%
\providecommand \selectlanguage [0]{\@gobble}%
\providecommand \bibinfo  [0]{\@secondoftwo}%
\providecommand \bibfield  [0]{\@secondoftwo}%
\providecommand \translation [1]{[#1]}%
\providecommand \BibitemOpen [0]{}%
\providecommand \bibitemStop [0]{}%
\providecommand \bibitemNoStop [0]{.\EOS\space}%
\providecommand \EOS [0]{\spacefactor3000\relax}%
\providecommand \BibitemShut  [1]{\csname bibitem#1\endcsname}%
\let\auto@bib@innerbib\@empty
%</preamble>
\bibitem [{\citenamefont {Emery}\ and\ \citenamefont
  {Kivelson}(1998)}]{Kivelson}%
  \BibitemOpen
  \bibfield  {author} {\bibinfo {author} {\bibfnamefont {J.~V.}\ \bibnamefont
  {Emery}}\ and\ \bibinfo {author} {\bibfnamefont {S.~A.}\ \bibnamefont
  {Kivelson}},\ }\href@noop {} {\bibfield  {journal} {\bibinfo  {journal}
  {Nature}\ }\textbf {\bibinfo {volume} {374}},\ \bibinfo {pages} {434}
  (\bibinfo {year} {1998})}\BibitemShut {NoStop}%
\bibitem [{\citenamefont {Engelbrecht}\ \emph {et~al.}(1998)\citenamefont
  {Engelbrecht}, \citenamefont {Nazarenko}, \citenamefont {Randeria},\ and\
  \citenamefont {Dagotto}}]{Engelbrecht}%
  \BibitemOpen
  \bibfield  {author} {\bibinfo {author} {\bibfnamefont {J.~R.}\ \bibnamefont
  {Engelbrecht}}, \bibinfo {author} {\bibfnamefont {A.}~\bibnamefont
  {Nazarenko}}, \bibinfo {author} {\bibfnamefont {M.}~\bibnamefont {Randeria}},
  \ and\ \bibinfo {author} {\bibfnamefont {E.}~\bibnamefont {Dagotto}},\
  }\href@noop {} {\bibfield  {journal} {\bibinfo  {journal} {Phys. Rev. B}\
  }\textbf {\bibinfo {volume} {57}},\ \bibinfo {pages} {13406} (\bibinfo {year}
  {1998})}\BibitemShut {NoStop}%
\bibitem [{\citenamefont {Simon}\ and\ \citenamefont {Varma}(2002)}]{Varma}%
  \BibitemOpen
  \bibfield  {author} {\bibinfo {author} {\bibfnamefont {M.~E.}\ \bibnamefont
  {Simon}}\ and\ \bibinfo {author} {\bibfnamefont {C.~M.}\ \bibnamefont
  {Varma}},\ }\href@noop {} {\bibfield  {journal} {\bibinfo  {journal} {Phys.
  Rev. Lett.}\ }\textbf {\bibinfo {volume} {89}},\ \bibinfo {pages} {247003}
  (\bibinfo {year} {2002})}\BibitemShut {NoStop}%
\bibitem [{\citenamefont {Deustcher}(1999)}]{Deustcher}%
  \BibitemOpen
  \bibfield  {author} {\bibinfo {author} {\bibfnamefont {G.}~\bibnamefont
  {Deustcher}},\ }\href@noop {} {\bibfield  {journal} {\bibinfo  {journal}
  {Nature}\ }\textbf {\bibinfo {volume} {397}},\ \bibinfo {pages} {410}
  (\bibinfo {year} {1999})}\BibitemShut {NoStop}%
\bibitem [{\citenamefont {Panagopoulos}\ \emph {et~al.}(1998)\citenamefont
  {Panagopoulos}, \citenamefont {Cooper},\ and\ \citenamefont
  {Xiang}}]{Panagopoulos}%
  \BibitemOpen
  \bibfield  {author} {\bibinfo {author} {\bibfnamefont {C.}~\bibnamefont
  {Panagopoulos}}, \bibinfo {author} {\bibfnamefont {J.~R.}\ \bibnamefont
  {Cooper}}, \ and\ \bibinfo {author} {\bibfnamefont {T.}~\bibnamefont
  {Xiang}},\ }\href@noop {} {\bibfield  {journal} {\bibinfo  {journal} {Phys.
  Rev. B}\ }\textbf {\bibinfo {volume} {57}},\ \bibinfo {pages} {13422}
  (\bibinfo {year} {1998})}\BibitemShut {NoStop}%
\bibitem [{\citenamefont {Opel}\ \emph {et~al.}(2000)\citenamefont {Opel},
  \citenamefont {Nemetschek}, \citenamefont {Hoffmann}, \citenamefont
  {Philipp}, \citenamefont {M\textrm{\"{u}}ller}, \citenamefont {Hackl},
  \citenamefont {T\textrm{\"{u}}tt\textrm{\"{o}}}, \citenamefont {Erb},
  \citenamefont {Revaz}, \citenamefont {Walker}, \citenamefont {Berger},\ and\
  \citenamefont {Forr\textrm{\'{o}}}}]{Opel}%
  \BibitemOpen
  \bibfield  {author} {\bibinfo {author} {\bibfnamefont {M.}~\bibnamefont
  {Opel}}, \bibinfo {author} {\bibfnamefont {R.}~\bibnamefont {Nemetschek}},
  \bibinfo {author} {\bibfnamefont {C.}~\bibnamefont {Hoffmann}}, \bibinfo
  {author} {\bibfnamefont {R.}~\bibnamefont {Philipp}}, \bibinfo {author}
  {\bibfnamefont {P.~F.}\ \bibnamefont {M\textrm{\"{u}}ller}}, \bibinfo
  {author} {\bibfnamefont {R.}~\bibnamefont {Hackl}}, \bibinfo {author}
  {\bibfnamefont {I.}~\bibnamefont {T\textrm{\"{u}}tt\textrm{\"{o}}}}, \bibinfo
  {author} {\bibfnamefont {A.}~\bibnamefont {Erb}}, \bibinfo {author}
  {\bibfnamefont {B.}~\bibnamefont {Revaz}}, \bibinfo {author} {\bibfnamefont
  {E.}~\bibnamefont {Walker}}, \bibinfo {author} {\bibfnamefont
  {H.}~\bibnamefont {Berger}}, \ and\ \bibinfo {author} {\bibfnamefont
  {L.}~\bibnamefont {Forr\textrm{\'{o}}}},\ }\href@noop {} {\bibfield
  {journal} {\bibinfo  {journal} {Phys. Rev. B}\ }\textbf {\bibinfo {volume}
  {61}},\ \bibinfo {pages} {9752} (\bibinfo {year} {2000})}\BibitemShut
  {NoStop}%
\bibitem [{\citenamefont {Tacon}\ \emph {et~al.}(2006)\citenamefont {Tacon},
  \citenamefont {Sacuto}, \citenamefont {Georges}, \citenamefont {Kotliar},
  \citenamefont {Gallais}, \citenamefont {Colson},\ and\ \citenamefont
  {Forget}}]{Tacon}%
  \BibitemOpen
  \bibfield  {author} {\bibinfo {author} {\bibfnamefont {M.~L.}\ \bibnamefont
  {Tacon}}, \bibinfo {author} {\bibfnamefont {A.}~\bibnamefont {Sacuto}},
  \bibinfo {author} {\bibfnamefont {A.}~\bibnamefont {Georges}}, \bibinfo
  {author} {\bibfnamefont {G.}~\bibnamefont {Kotliar}}, \bibinfo {author}
  {\bibfnamefont {Y.}~\bibnamefont {Gallais}}, \bibinfo {author} {\bibfnamefont
  {D.}~\bibnamefont {Colson}}, \ and\ \bibinfo {author} {\bibfnamefont
  {A.}~\bibnamefont {Forget}},\ }\href@noop {} {\bibfield  {journal} {\bibinfo
  {journal} {Nature Physics}\ }\textbf {\bibinfo {volume} {2}},\ \bibinfo
  {pages} {537} (\bibinfo {year} {2006})}\BibitemShut {NoStop}%
\bibitem [{\citenamefont {Tanaka}\ \emph {et~al.}(2006)\citenamefont {Tanaka},
  \citenamefont {Lee}, \citenamefont {Lu}, \citenamefont {Fujimori},
  \citenamefont {Fujii}, \citenamefont {Risdiana}, \citenamefont {Terasaki},
  \citenamefont {Scalapino}, \citenamefont {Devereaux}, \citenamefont
  {Hussain},\ and\ \citenamefont {Shen}}]{Tanaka}%
  \BibitemOpen
  \bibfield  {author} {\bibinfo {author} {\bibfnamefont {K.}~\bibnamefont
  {Tanaka}}, \bibinfo {author} {\bibfnamefont {W.~S.}\ \bibnamefont {Lee}},
  \bibinfo {author} {\bibfnamefont {D.~H.}\ \bibnamefont {Lu}}, \bibinfo
  {author} {\bibfnamefont {A.}~\bibnamefont {Fujimori}}, \bibinfo {author}
  {\bibfnamefont {T.}~\bibnamefont {Fujii}}, \bibinfo {author} {\bibnamefont
  {Risdiana}}, \bibinfo {author} {\bibfnamefont {I.}~\bibnamefont {Terasaki}},
  \bibinfo {author} {\bibfnamefont {D.~J.}\ \bibnamefont {Scalapino}}, \bibinfo
  {author} {\bibfnamefont {T.~P.}\ \bibnamefont {Devereaux}}, \bibinfo {author}
  {\bibfnamefont {Z.}~\bibnamefont {Hussain}}, \ and\ \bibinfo {author}
  {\bibfnamefont {Z.-X.}\ \bibnamefont {Shen}},\ }\href@noop {} {\bibfield
  {journal} {\bibinfo  {journal} {Science}\ }\textbf {\bibinfo {volume}
  {314}},\ \bibinfo {pages} {1910} (\bibinfo {year} {2006})}\BibitemShut
  {NoStop}%
\bibitem [{\citenamefont {Lee}\ \emph {et~al.}(2007)\citenamefont {Lee},
  \citenamefont {Vishik}, \citenamefont {Tanaka}, \citenamefont {Lu},
  \citenamefont {Sasagawa}, \citenamefont {Nagaosa}, \citenamefont {Devereaux},
  \citenamefont {Hussain},\ and\ \citenamefont {Shen}}]{Lee}%
  \BibitemOpen
  \bibfield  {author} {\bibinfo {author} {\bibfnamefont {W.~S.}\ \bibnamefont
  {Lee}}, \bibinfo {author} {\bibfnamefont {I.~M.}\ \bibnamefont {Vishik}},
  \bibinfo {author} {\bibfnamefont {K.}~\bibnamefont {Tanaka}}, \bibinfo
  {author} {\bibfnamefont {D.~H.}\ \bibnamefont {Lu}}, \bibinfo {author}
  {\bibfnamefont {T.}~\bibnamefont {Sasagawa}}, \bibinfo {author}
  {\bibfnamefont {N.}~\bibnamefont {Nagaosa}}, \bibinfo {author} {\bibfnamefont
  {T.~P.}\ \bibnamefont {Devereaux}}, \bibinfo {author} {\bibfnamefont
  {Z.}~\bibnamefont {Hussain}}, \ and\ \bibinfo {author} {\bibfnamefont
  {Z.-X.}\ \bibnamefont {Shen}},\ }\href@noop {} {\bibfield  {journal}
  {\bibinfo  {journal} {Nature}\ }\textbf {\bibinfo {volume} {450}},\ \bibinfo
  {pages} {81} (\bibinfo {year} {2007})}\BibitemShut {NoStop}%
\bibitem [{\citenamefont {Kondo}\ \emph {et~al.}(2007)\citenamefont {Kondo},
  \citenamefont {Takeuchi}, \citenamefont {Kaminski}, \citenamefont {Tsuda},\
  and\ \citenamefont {Shin}}]{Kondo}%
  \BibitemOpen
  \bibfield  {author} {\bibinfo {author} {\bibfnamefont {T.}~\bibnamefont
  {Kondo}}, \bibinfo {author} {\bibfnamefont {T.}~\bibnamefont {Takeuchi}},
  \bibinfo {author} {\bibfnamefont {A.}~\bibnamefont {Kaminski}}, \bibinfo
  {author} {\bibfnamefont {S.}~\bibnamefont {Tsuda}}, \ and\ \bibinfo {author}
  {\bibfnamefont {S.}~\bibnamefont {Shin}},\ }\href@noop {} {\bibfield
  {journal} {\bibinfo  {journal} {Phys. Rev. Lett.}\ }\textbf {\bibinfo
  {volume} {98}},\ \bibinfo {pages} {267004} (\bibinfo {year}
  {2007})}\BibitemShut {NoStop}%
\bibitem [{\citenamefont {Terashima}\ \emph {et~al.}(2007)\citenamefont
  {Terashima}, \citenamefont {Matsui}, \citenamefont {Sato}, \citenamefont
  {Takahashi}, \citenamefont {Kofu},\ and\ \citenamefont {Hirota}}]{Terashima}%
  \BibitemOpen
  \bibfield  {author} {\bibinfo {author} {\bibfnamefont {K.}~\bibnamefont
  {Terashima}}, \bibinfo {author} {\bibfnamefont {H.}~\bibnamefont {Matsui}},
  \bibinfo {author} {\bibfnamefont {T.}~\bibnamefont {Sato}}, \bibinfo {author}
  {\bibfnamefont {T.}~\bibnamefont {Takahashi}}, \bibinfo {author}
  {\bibfnamefont {M.}~\bibnamefont {Kofu}}, \ and\ \bibinfo {author}
  {\bibfnamefont {K.}~\bibnamefont {Hirota}},\ }\href@noop {} {\bibfield
  {journal} {\bibinfo  {journal} {Phys. Rev. Lett.}\ }\textbf {\bibinfo
  {volume} {99}},\ \bibinfo {pages} {017003} (\bibinfo {year}
  {2007})}\BibitemShut {NoStop}%
\bibitem [{\citenamefont {Hashimoto}\ \emph {et~al.}(2007)\citenamefont
  {Hashimoto}, \citenamefont {Yoshida}, \citenamefont {Tanaka}, \citenamefont
  {Fujimori}, \citenamefont {Okusawa}, \citenamefont {Wakimoto}, \citenamefont
  {Yamada}, \citenamefont {Kakeshita}, \citenamefont {Eisaki},\ and\
  \citenamefont {Uchida}}]{Hashimoto}%
  \BibitemOpen
  \bibfield  {author} {\bibinfo {author} {\bibfnamefont {M.}~\bibnamefont
  {Hashimoto}}, \bibinfo {author} {\bibfnamefont {T.}~\bibnamefont {Yoshida}},
  \bibinfo {author} {\bibfnamefont {K.}~\bibnamefont {Tanaka}}, \bibinfo
  {author} {\bibfnamefont {A.}~\bibnamefont {Fujimori}}, \bibinfo {author}
  {\bibfnamefont {M.}~\bibnamefont {Okusawa}}, \bibinfo {author} {\bibfnamefont
  {S.}~\bibnamefont {Wakimoto}}, \bibinfo {author} {\bibfnamefont
  {K.}~\bibnamefont {Yamada}}, \bibinfo {author} {\bibfnamefont
  {T.}~\bibnamefont {Kakeshita}}, \bibinfo {author} {\bibfnamefont
  {H.}~\bibnamefont {Eisaki}}, \ and\ \bibinfo {author} {\bibfnamefont
  {S.}~\bibnamefont {Uchida}},\ }\href@noop {} {\bibfield  {journal} {\bibinfo
  {journal} {Phys. Rev. B}\ }\textbf {\bibinfo {volume} {75}},\ \bibinfo
  {pages} {140503(R)} (\bibinfo {year} {2007})}\BibitemShut {NoStop}%
\bibitem [{\citenamefont {Yoshida}\ \emph {et~al.}(2009)\citenamefont
  {Yoshida}, \citenamefont {Hashimoto}, \citenamefont {Ideta}, \citenamefont
  {Fujimori}, \citenamefont {Tanaka}, \citenamefont {Mannella}, \citenamefont
  {Hussain}, \citenamefont {Shen}, \citenamefont {Kubota}, \citenamefont {Ono},
  \citenamefont {Komiya}, \citenamefont {Ando}, \citenamefont {Eisaki},\ and\
  \citenamefont {Uchida}}]{Yoshida}%
  \BibitemOpen
  \bibfield  {author} {\bibinfo {author} {\bibfnamefont {T.}~\bibnamefont
  {Yoshida}}, \bibinfo {author} {\bibfnamefont {M.}~\bibnamefont {Hashimoto}},
  \bibinfo {author} {\bibfnamefont {S.}~\bibnamefont {Ideta}}, \bibinfo
  {author} {\bibfnamefont {A.}~\bibnamefont {Fujimori}}, \bibinfo {author}
  {\bibfnamefont {K.}~\bibnamefont {Tanaka}}, \bibinfo {author} {\bibfnamefont
  {N.}~\bibnamefont {Mannella}}, \bibinfo {author} {\bibfnamefont
  {Z.}~\bibnamefont {Hussain}}, \bibinfo {author} {\bibfnamefont {Z.-X.}\
  \bibnamefont {Shen}}, \bibinfo {author} {\bibfnamefont {M.}~\bibnamefont
  {Kubota}}, \bibinfo {author} {\bibfnamefont {K.}~\bibnamefont {Ono}},
  \bibinfo {author} {\bibfnamefont {S.}~\bibnamefont {Komiya}}, \bibinfo
  {author} {\bibfnamefont {Y.}~\bibnamefont {Ando}}, \bibinfo {author}
  {\bibfnamefont {H.}~\bibnamefont {Eisaki}}, \ and\ \bibinfo {author}
  {\bibfnamefont {S.}~\bibnamefont {Uchida}},\ }\href {\doibase
  10.1103/PhysRevLett.103.037004} {\bibfield  {journal} {\bibinfo  {journal}
  {Phys. Rev. Lett.}\ }\textbf {\bibinfo {volume} {103}},\ \bibinfo {pages}
  {037004} (\bibinfo {year} {2009})}\BibitemShut {NoStop}%
\bibitem [{\citenamefont {Meng}\ \emph {et~al.}(2009)\citenamefont {Meng},
  \citenamefont {Zhang}, \citenamefont {Liu}, \citenamefont {Zhao},
  \citenamefont {Liu}, \citenamefont {Jia}, \citenamefont {Lu}, \citenamefont
  {Dong}, \citenamefont {Wang}, \citenamefont {Zhang}, \citenamefont {Zhou},
  \citenamefont {Zhu}, \citenamefont {Wang}, \citenamefont {Zhao},
  \citenamefont {Xu}, \citenamefont {Chen},\ and\ \citenamefont {Zhou}}]{Meng}%
  \BibitemOpen
  \bibfield  {author} {\bibinfo {author} {\bibfnamefont {J.}~\bibnamefont
  {Meng}}, \bibinfo {author} {\bibfnamefont {W.}~\bibnamefont {Zhang}},
  \bibinfo {author} {\bibfnamefont {G.}~\bibnamefont {Liu}}, \bibinfo {author}
  {\bibfnamefont {L.}~\bibnamefont {Zhao}}, \bibinfo {author} {\bibfnamefont
  {H.}~\bibnamefont {Liu}}, \bibinfo {author} {\bibfnamefont {X.}~\bibnamefont
  {Jia}}, \bibinfo {author} {\bibfnamefont {W.}~\bibnamefont {Lu}}, \bibinfo
  {author} {\bibfnamefont {X.}~\bibnamefont {Dong}}, \bibinfo {author}
  {\bibfnamefont {G.}~\bibnamefont {Wang}}, \bibinfo {author} {\bibfnamefont
  {H.}~\bibnamefont {Zhang}}, \bibinfo {author} {\bibfnamefont
  {Y.}~\bibnamefont {Zhou}}, \bibinfo {author} {\bibfnamefont {Y.}~\bibnamefont
  {Zhu}}, \bibinfo {author} {\bibfnamefont {X.}~\bibnamefont {Wang}}, \bibinfo
  {author} {\bibfnamefont {Z.}~\bibnamefont {Zhao}}, \bibinfo {author}
  {\bibfnamefont {Z.}~\bibnamefont {Xu}}, \bibinfo {author} {\bibfnamefont
  {C.}~\bibnamefont {Chen}}, \ and\ \bibinfo {author} {\bibfnamefont {X.~J.}\
  \bibnamefont {Zhou}},\ }\href {\doibase 10.1103/PhysRevB.79.024514}
  {\bibfield  {journal} {\bibinfo  {journal} {Phys. Rev. B}\ }\textbf {\bibinfo
  {volume} {79}},\ \bibinfo {pages} {024514} (\bibinfo {year}
  {2009})}\BibitemShut {NoStop}%
\bibitem [{\citenamefont {Shi}\ \emph {et~al.}(2009)\citenamefont {Shi},
  \citenamefont {Bendounan}, \citenamefont {Razzoli}, \citenamefont
  {Rosenkranz}, \citenamefont {Norman}, \citenamefont {Campuzano},
  \citenamefont {Chang}, \citenamefont {Mansson}, \citenamefont {Sassa},
  \citenamefont {Claesson}, \citenamefont {Tjernberg}, \citenamefont {Patthey},
  \citenamefont {Momono}, \citenamefont {Oda}, \citenamefont {Ido},
  \citenamefont {Guerrero}, \citenamefont {Mudry},\ and\ \citenamefont
  {Mesot}}]{Shi}%
  \BibitemOpen
  \bibfield  {author} {\bibinfo {author} {\bibfnamefont {M.}~\bibnamefont
  {Shi}}, \bibinfo {author} {\bibfnamefont {A.}~\bibnamefont {Bendounan}},
  \bibinfo {author} {\bibfnamefont {E.}~\bibnamefont {Razzoli}}, \bibinfo
  {author} {\bibfnamefont {S.}~\bibnamefont {Rosenkranz}}, \bibinfo {author}
  {\bibfnamefont {M.~R.}\ \bibnamefont {Norman}}, \bibinfo {author}
  {\bibfnamefont {J.~C.}\ \bibnamefont {Campuzano}}, \bibinfo {author}
  {\bibfnamefont {J.}~\bibnamefont {Chang}}, \bibinfo {author} {\bibfnamefont
  {M.}~\bibnamefont {Mansson}}, \bibinfo {author} {\bibfnamefont
  {Y.}~\bibnamefont {Sassa}}, \bibinfo {author} {\bibfnamefont
  {T.}~\bibnamefont {Claesson}}, \bibinfo {author} {\bibfnamefont
  {O.}~\bibnamefont {Tjernberg}}, \bibinfo {author} {\bibfnamefont
  {L.}~\bibnamefont {Patthey}}, \bibinfo {author} {\bibfnamefont
  {N.}~\bibnamefont {Momono}}, \bibinfo {author} {\bibfnamefont
  {M.}~\bibnamefont {Oda}}, \bibinfo {author} {\bibfnamefont {M.}~\bibnamefont
  {Ido}}, \bibinfo {author} {\bibfnamefont {S.}~\bibnamefont {Guerrero}},
  \bibinfo {author} {\bibfnamefont {C.}~\bibnamefont {Mudry}}, \ and\ \bibinfo
  {author} {\bibfnamefont {J.}~\bibnamefont {Mesot}},\ }\href {\doibase
  10.1209/0295-5075/88/27008} {\bibfield  {journal} {\bibinfo  {journal}
  {Europhys. Lett.}\ }\textbf {\bibinfo {volume} {88}},\ \bibinfo {pages}
  {27008} (\bibinfo {year} {2009})}\BibitemShut {NoStop}%
\bibitem [{\citenamefont {Boyer}\ \emph {et~al.}(2007)\citenamefont {Boyer},
  \citenamefont {Wise}, \citenamefont {Chatterjee}, \citenamefont {Yi},
  \citenamefont {Kondo}, \citenamefont {Takeuchi}, \citenamefont {Ikuta},\ and\
  \citenamefont {Hudson}}]{Boyer}%
  \BibitemOpen
  \bibfield  {author} {\bibinfo {author} {\bibfnamefont {M.~C.}\ \bibnamefont
  {Boyer}}, \bibinfo {author} {\bibfnamefont {W.~D.}\ \bibnamefont {Wise}},
  \bibinfo {author} {\bibfnamefont {K.}~\bibnamefont {Chatterjee}}, \bibinfo
  {author} {\bibfnamefont {M.}~\bibnamefont {Yi}}, \bibinfo {author}
  {\bibfnamefont {T.}~\bibnamefont {Kondo}}, \bibinfo {author} {\bibfnamefont
  {T.}~\bibnamefont {Takeuchi}}, \bibinfo {author} {\bibfnamefont
  {H.}~\bibnamefont {Ikuta}}, \ and\ \bibinfo {author} {\bibfnamefont {E.~W.}\
  \bibnamefont {Hudson}},\ }\href {\doibase 10.1038/nphys725} {\bibfield
  {journal} {\bibinfo  {journal} {Nature Physics}\ }\textbf {\bibinfo {volume}
  {3}},\ \bibinfo {pages} {802} (\bibinfo {year} {2007})}\BibitemShut {NoStop}%
\bibitem [{\citenamefont {Ma}\ \emph {et~al.}(2008)\citenamefont {Ma},
  \citenamefont {Pan}, \citenamefont {Niestemski}, \citenamefont {Neupane},
  \citenamefont {Xu}, \citenamefont {Richard}, \citenamefont {Nakayama},
  \citenamefont {Sato}, \citenamefont {Takahashi}, \citenamefont {Luo},
  \citenamefont {Fang}, \citenamefont {Wen}, \citenamefont {Wang},
  \citenamefont {Ding},\ and\ \citenamefont {Madhavan}}]{Ma}%
  \BibitemOpen
  \bibfield  {author} {\bibinfo {author} {\bibfnamefont {J.-H.}\ \bibnamefont
  {Ma}}, \bibinfo {author} {\bibfnamefont {Z.-H.}\ \bibnamefont {Pan}},
  \bibinfo {author} {\bibfnamefont {F.~C.}\ \bibnamefont {Niestemski}},
  \bibinfo {author} {\bibfnamefont {M.}~\bibnamefont {Neupane}}, \bibinfo
  {author} {\bibfnamefont {Y.-M.}\ \bibnamefont {Xu}}, \bibinfo {author}
  {\bibfnamefont {P.}~\bibnamefont {Richard}}, \bibinfo {author} {\bibfnamefont
  {K.}~\bibnamefont {Nakayama}}, \bibinfo {author} {\bibfnamefont
  {T.}~\bibnamefont {Sato}}, \bibinfo {author} {\bibfnamefont {T.}~\bibnamefont
  {Takahashi}}, \bibinfo {author} {\bibfnamefont {H.-Q.}\ \bibnamefont {Luo}},
  \bibinfo {author} {\bibfnamefont {L.}~\bibnamefont {Fang}}, \bibinfo {author}
  {\bibfnamefont {H.-H.}\ \bibnamefont {Wen}}, \bibinfo {author} {\bibfnamefont
  {Z.}~\bibnamefont {Wang}}, \bibinfo {author} {\bibfnamefont {H.}~\bibnamefont
  {Ding}}, \ and\ \bibinfo {author} {\bibfnamefont {V.}~\bibnamefont
  {Madhavan}},\ }\href {\doibase 10.1103/PhysRevLett.101.207002} {\bibfield
  {journal} {\bibinfo  {journal} {Phys. Rev. Lett.}\ }\textbf {\bibinfo
  {volume} {101}},\ \bibinfo {pages} {207002} (\bibinfo {year}
  {2008})}\BibitemShut {NoStop}%
\bibitem [{\citenamefont {Ding}\ \emph {et~al.}(2001)\citenamefont {Ding},
  \citenamefont {Engelbrecht}, \citenamefont {Wang}, \citenamefont {Campuzano},
  \citenamefont {Wang}, \citenamefont {Yang}, \citenamefont {Rogan},
  \citenamefont {Takahashi}, \citenamefont {Kadowaki},\ and\ \citenamefont
  {Hinks}}]{Ding}%
  \BibitemOpen
  \bibfield  {author} {\bibinfo {author} {\bibfnamefont {H.}~\bibnamefont
  {Ding}}, \bibinfo {author} {\bibfnamefont {J.~R.}\ \bibnamefont
  {Engelbrecht}}, \bibinfo {author} {\bibfnamefont {Z.}~\bibnamefont {Wang}},
  \bibinfo {author} {\bibfnamefont {J.~C.}\ \bibnamefont {Campuzano}}, \bibinfo
  {author} {\bibfnamefont {S.-C.}\ \bibnamefont {Wang}}, \bibinfo {author}
  {\bibfnamefont {H.-B.}\ \bibnamefont {Yang}}, \bibinfo {author}
  {\bibfnamefont {R.}~\bibnamefont {Rogan}}, \bibinfo {author} {\bibfnamefont
  {T.}~\bibnamefont {Takahashi}}, \bibinfo {author} {\bibfnamefont
  {K.}~\bibnamefont {Kadowaki}}, \ and\ \bibinfo {author} {\bibfnamefont
  {D.~G.}\ \bibnamefont {Hinks}},\ }\href@noop {} {\bibfield  {journal}
  {\bibinfo  {journal} {Phys. Rev. Lett.}\ }\textbf {\bibinfo {volume} {87}},\
  \bibinfo {pages} {227001} (\bibinfo {year} {2001})}\BibitemShut {NoStop}%
\bibitem [{\citenamefont {McMillan}\ and\ \citenamefont
  {Rowell}(1965)}]{McMillan}%
  \BibitemOpen
  \bibfield  {author} {\bibinfo {author} {\bibfnamefont {W.~L.}\ \bibnamefont
  {McMillan}}\ and\ \bibinfo {author} {\bibfnamefont {J.~M.}\ \bibnamefont
  {Rowell}},\ }\href {\doibase 10.1103/PhysRevLett.14.108} {\bibfield
  {journal} {\bibinfo  {journal} {Phys. Rev. Lett.}\ }\textbf {\bibinfo
  {volume} {14}},\ \bibinfo {pages} {108} (\bibinfo {year} {1965})}\BibitemShut
  {NoStop}%
\bibitem [{\citenamefont {Ideta}\ \emph {et~al.}(2012)\citenamefont {Ideta},
  \citenamefont {Yoshida}, \citenamefont {Fujimori}, \citenamefont {Anzai},
  \citenamefont {Fujita}, \citenamefont {Ino}, \citenamefont {Arita},
  \citenamefont {Namatame}, \citenamefont {Taniguchi}, \citenamefont {Shen},
  \citenamefont {Takashima}, \citenamefont {Kojima},\ and\ \citenamefont
  {Uchida}}]{Ideta}%
  \BibitemOpen
  \bibfield  {author} {\bibinfo {author} {\bibfnamefont {S.-i.}\ \bibnamefont
  {Ideta}}, \bibinfo {author} {\bibfnamefont {T.}~\bibnamefont {Yoshida}},
  \bibinfo {author} {\bibfnamefont {A.}~\bibnamefont {Fujimori}}, \bibinfo
  {author} {\bibfnamefont {H.}~\bibnamefont {Anzai}}, \bibinfo {author}
  {\bibfnamefont {T.}~\bibnamefont {Fujita}}, \bibinfo {author} {\bibfnamefont
  {A.}~\bibnamefont {Ino}}, \bibinfo {author} {\bibfnamefont {M.}~\bibnamefont
  {Arita}}, \bibinfo {author} {\bibfnamefont {H.}~\bibnamefont {Namatame}},
  \bibinfo {author} {\bibfnamefont {M.}~\bibnamefont {Taniguchi}}, \bibinfo
  {author} {\bibfnamefont {Z.-X.}\ \bibnamefont {Shen}}, \bibinfo {author}
  {\bibfnamefont {K.}~\bibnamefont {Takashima}}, \bibinfo {author}
  {\bibfnamefont {K.}~\bibnamefont {Kojima}}, \ and\ \bibinfo {author}
  {\bibfnamefont {S.-i.}\ \bibnamefont {Uchida}},\ }\href@noop {} {\bibfield
  {journal} {\bibinfo  {journal} {Phys. Rev. B}\ }\textbf {\bibinfo {volume}
  {85}},\ \bibinfo {pages} {104515} (\bibinfo {year} {2012})}\BibitemShut
  {NoStop}%
\bibitem [{\citenamefont {Yoshida}\ \emph {et~al.}(2012)\citenamefont
  {Yoshida}, \citenamefont {Hashimoto}, \citenamefont {Vishik}, \citenamefont
  {Shen},\ and\ \citenamefont {Fujimori}}]{PGreview}%
  \BibitemOpen
  \bibfield  {author} {\bibinfo {author} {\bibfnamefont {T.}~\bibnamefont
  {Yoshida}}, \bibinfo {author} {\bibfnamefont {M.}~\bibnamefont {Hashimoto}},
  \bibinfo {author} {\bibfnamefont {I.~M.}\ \bibnamefont {Vishik}}, \bibinfo
  {author} {\bibfnamefont {Z.-X.}\ \bibnamefont {Shen}}, \ and\ \bibinfo
  {author} {\bibfnamefont {A.}~\bibnamefont {Fujimori}},\ }\href@noop {}
  {\bibfield  {journal} {\bibinfo  {journal} {Journal of the Physical Society
  of Japan}\ }\textbf {\bibinfo {volume} {81}},\ \bibinfo {pages} {011006}
  (\bibinfo {year} {2012})}\BibitemShut {NoStop}%
\bibitem [{\citenamefont {Kurosawa}\ \emph {et~al.}(2010)\citenamefont
  {Kurosawa}, \citenamefont {Yoneyama}, \citenamefont {Takano}, \citenamefont
  {Hagiwara}, \citenamefont {Inoue}, \citenamefont {Hagiwara}, \citenamefont
  {Kurusu}, \citenamefont {Takeyama}, \citenamefont {Momono}, \citenamefont
  {Oda},\ and\ \citenamefont {Ido}}]{Kurosawa}%
  \BibitemOpen
  \bibfield  {author} {\bibinfo {author} {\bibfnamefont {T.}~\bibnamefont
  {Kurosawa}}, \bibinfo {author} {\bibfnamefont {T.}~\bibnamefont {Yoneyama}},
  \bibinfo {author} {\bibfnamefont {Y.}~\bibnamefont {Takano}}, \bibinfo
  {author} {\bibfnamefont {M.}~\bibnamefont {Hagiwara}}, \bibinfo {author}
  {\bibfnamefont {R.}~\bibnamefont {Inoue}}, \bibinfo {author} {\bibfnamefont
  {N.}~\bibnamefont {Hagiwara}}, \bibinfo {author} {\bibfnamefont
  {K.}~\bibnamefont {Kurusu}}, \bibinfo {author} {\bibfnamefont
  {K.}~\bibnamefont {Takeyama}}, \bibinfo {author} {\bibfnamefont
  {N.}~\bibnamefont {Momono}}, \bibinfo {author} {\bibfnamefont
  {M.}~\bibnamefont {Oda}}, \ and\ \bibinfo {author} {\bibfnamefont
  {M.}~\bibnamefont {Ido}},\ }\href {\doibase 10.1103/PhysRevB.81.094519}
  {\bibfield  {journal} {\bibinfo  {journal} {Phys. Rev. B}\ }\textbf {\bibinfo
  {volume} {81}},\ \bibinfo {pages} {094519} (\bibinfo {year}
  {2010})}\BibitemShut {NoStop}%
\bibitem [{\citenamefont {Yang}\ \emph {et~al.}(2006)\citenamefont {Yang},
  \citenamefont {Rice},\ and\ \citenamefont {Zhang}}]{YRZ}%
  \BibitemOpen
  \bibfield  {author} {\bibinfo {author} {\bibfnamefont {K.-Y.}\ \bibnamefont
  {Yang}}, \bibinfo {author} {\bibfnamefont {T.~M.}\ \bibnamefont {Rice}}, \
  and\ \bibinfo {author} {\bibfnamefont {F.-C.}\ \bibnamefont {Zhang}},\ }\href
  {\doibase 10.1103/PhysRevB.73.174501} {\bibfield  {journal} {\bibinfo
  {journal} {Phys. Rev. B}\ }\textbf {\bibinfo {volume} {73}},\ \bibinfo
  {pages} {174501} (\bibinfo {year} {2006})}\BibitemShut {NoStop}%
\bibitem [{\citenamefont {Niestemski}\ and\ \citenamefont {Wang}(2009)}]{Ren}%
  \BibitemOpen
  \bibfield  {author} {\bibinfo {author} {\bibfnamefont {L.~R.}\ \bibnamefont
  {Niestemski}}\ and\ \bibinfo {author} {\bibfnamefont {Z.}~\bibnamefont
  {Wang}},\ }\href {\doibase 10.1103/PhysRevLett.102.107001} {\bibfield
  {journal} {\bibinfo  {journal} {Phys. Rev. Lett.}\ }\textbf {\bibinfo
  {volume} {102}},\ \bibinfo {pages} {107001} (\bibinfo {year}
  {2009})}\BibitemShut {NoStop}%
\bibitem [{\citenamefont {Das}\ \emph {et~al.}(2008)\citenamefont {Das},
  \citenamefont {Markiewicz},\ and\ \citenamefont {Bansil}}]{Das}%
  \BibitemOpen
  \bibfield  {author} {\bibinfo {author} {\bibfnamefont {T.}~\bibnamefont
  {Das}}, \bibinfo {author} {\bibfnamefont {R.~S.}\ \bibnamefont {Markiewicz}},
  \ and\ \bibinfo {author} {\bibfnamefont {A.}~\bibnamefont {Bansil}},\ }\href
  {\doibase 10.1103/PhysRevB.77.134516} {\bibfield  {journal} {\bibinfo
  {journal} {Phys. Rev. B}\ }\textbf {\bibinfo {volume} {77}},\ \bibinfo
  {pages} {134516} (\bibinfo {year} {2008})}\BibitemShut {NoStop}%
\bibitem [{\citenamefont {Hashimoto}\ \emph {et~al.}(2010)\citenamefont
  {Hashimoto}, \citenamefont {He}, \citenamefont {Tanaka}, \citenamefont
  {Testaud}, \citenamefont {Meevasana}, \citenamefont {Moore}, \citenamefont
  {Lu}, \citenamefont {Yao}, \citenamefont {Yoshida}, \citenamefont {Eisaki},
  \citenamefont {Devereaux}, \citenamefont {Hussain},\ and\ \citenamefont
  {Shen}}]{HashimotoNatPhys}%
  \BibitemOpen
  \bibfield  {author} {\bibinfo {author} {\bibfnamefont {M.}~\bibnamefont
  {Hashimoto}}, \bibinfo {author} {\bibfnamefont {R.-H.}\ \bibnamefont {He}},
  \bibinfo {author} {\bibfnamefont {K.}~\bibnamefont {Tanaka}}, \bibinfo
  {author} {\bibfnamefont {J.-P.}\ \bibnamefont {Testaud}}, \bibinfo {author}
  {\bibfnamefont {W.}~\bibnamefont {Meevasana}}, \bibinfo {author}
  {\bibfnamefont {R.~G.}\ \bibnamefont {Moore}}, \bibinfo {author}
  {\bibfnamefont {D.}~\bibnamefont {Lu}}, \bibinfo {author} {\bibfnamefont
  {H.}~\bibnamefont {Yao}}, \bibinfo {author} {\bibfnamefont {Y.}~\bibnamefont
  {Yoshida}}, \bibinfo {author} {\bibfnamefont {H.}~\bibnamefont {Eisaki}},
  \bibinfo {author} {\bibfnamefont {T.~P.}\ \bibnamefont {Devereaux}}, \bibinfo
  {author} {\bibfnamefont {Z.}~\bibnamefont {Hussain}}, \ and\ \bibinfo
  {author} {\bibfnamefont {Z.-X.}\ \bibnamefont {Shen}},\ }\href@noop {}
  {\bibfield  {journal} {\bibinfo  {journal} {Nature Physics}\ }\textbf
  {\bibinfo {volume} {6}},\ \bibinfo {pages} {414} (\bibinfo {year}
  {2010})}\BibitemShut {NoStop}%
\bibitem [{\citenamefont {Hanaguri}\ \emph {et~al.}(2004)\citenamefont
  {Hanaguri}, \citenamefont {Lupien}, \citenamefont {Kohsaka}, \citenamefont
  {Lee}, \citenamefont {Azuma}, \citenamefont {Takano}, \citenamefont
  {Takagi},\ and\ \citenamefont {Davis}}]{Hanaguri}%
  \BibitemOpen
  \bibfield  {author} {\bibinfo {author} {\bibfnamefont {T.}~\bibnamefont
  {Hanaguri}}, \bibinfo {author} {\bibfnamefont {C.}~\bibnamefont {Lupien}},
  \bibinfo {author} {\bibfnamefont {Y.}~\bibnamefont {Kohsaka}}, \bibinfo
  {author} {\bibfnamefont {D.-H.}\ \bibnamefont {Lee}}, \bibinfo {author}
  {\bibfnamefont {M.}~\bibnamefont {Azuma}}, \bibinfo {author} {\bibfnamefont
  {M.}~\bibnamefont {Takano}}, \bibinfo {author} {\bibfnamefont
  {H.}~\bibnamefont {Takagi}}, \ and\ \bibinfo {author} {\bibfnamefont {J.~C.}\
  \bibnamefont {Davis}},\ }\href@noop {} {\bibfield  {journal} {\bibinfo
  {journal} {Nature}\ }\textbf {\bibinfo {volume} {430}},\ \bibinfo {pages}
  {1001} (\bibinfo {year} {2004})}\BibitemShut {NoStop}%
\bibitem [{\citenamefont {Tranquada}\ \emph {et~al.}(1995)\citenamefont
  {Tranquada}, \citenamefont {Sternlieb}, \citenamefont {Axe}, \citenamefont
  {Nakamura},\ and\ \citenamefont {Uchida}}]{Tranquada}%
  \BibitemOpen
  \bibfield  {author} {\bibinfo {author} {\bibfnamefont {J.~M.}\ \bibnamefont
  {Tranquada}}, \bibinfo {author} {\bibfnamefont {B.~J.}\ \bibnamefont
  {Sternlieb}}, \bibinfo {author} {\bibfnamefont {J.~D.}\ \bibnamefont {Axe}},
  \bibinfo {author} {\bibfnamefont {Y.}~\bibnamefont {Nakamura}}, \ and\
  \bibinfo {author} {\bibfnamefont {S.}~\bibnamefont {Uchida}},\ }\href@noop {}
  {\bibfield  {journal} {\bibinfo  {journal} {Nature}\ }\textbf {\bibinfo
  {volume} {375}},\ \bibinfo {pages} {561} (\bibinfo {year}
  {1995})}\BibitemShut {NoStop}%
\end{thebibliography}%

\end{document}